\documentclass[preprint,preprintnumbers,aps,prl,showpacs]{revtex4}
\usepackage{graphicx}
\usepackage{bm}

\def\prl#1#2#3{{ Phys. Rev. Lett.} {\bf #1}, #2 (#3).}

\def\ibid#1#2#3{{ ibid.} {\bf #1}, #2 (#3).}
\def\pre#1#2#3{{  Phys. Rev. E.} {\bf #1}, #2 (#3).}
\def\pra#1#2#3{{ Phys. Rev. A} {\bf #1}, #2 (#3).}
\def\prb#1#2#3{{ Phys. Rev. B} {\bf #1}, #2 (#3).}
\def\jcp#1#2#3{{ J. Chem. Phys.} {\bf #1}, #2 (#3).}

\def\jpcb#1#2#3{{ J. Phys. Chem. B} {\bf #1}, #2 (#3).}

\def\nature#1#2#3{ { Nature} {\bf #1}, #2 (#3).}

\def\jpcm#1#2#3{ { J. Phys.: Condensed Matter } {\bf #1}, #2 (#3).}

\def\pccp#1#2#3{ { Phys. Chem. Chem. Phys.} {\bf #1}, #2 (#3).}

\def\expt#1{\langle #1\rangle}
\def\be{\begin{equation}}
\def\ee{\end{equation}}

\begin{document}
\voffset 0.5truein
\title{Entropy, Diffusivity and Structural Order in Liquids with
Water-like Anomalies}
\author{Ruchi Sharma}, 
\affiliation{Department of Chemistry, Indian Institute of Technology Delhi,
Hauz Khas, New Delhi: 110016, India.}
\author{Somendra Nath Chakraborty} 
\affiliation{Department of Chemistry, Indian Institute of Technology Delhi,
Hauz Khas, New Delhi: 110016, India.}
\author{Charusita  Chakravarty}
\email{charus@chemistry.iitd.ernet.in}
\affiliation{Department of Chemistry, Indian Institute of Technology Delhi, 
Hauz Khas, New Delhi: 110016, India.}

\begin{abstract}
The excess entropy, $S_{e}$, defined as the difference between 
the entropies of the liquid and the ideal gas under identical density 
and temperature conditions, is shown to be the critical quantity 
connecting the structural, diffusional and density anomalies in water-like
liquids.  Based on simulations of silica and the two-scale
ramp liquids, water-like density and diffusional anomalies can be seen
as consequences  of a characteristic non-monotonic density dependence of 
$S_{e}$.  The relationship between excess entropy, the order metrics and the 
structural anomaly can be understood 
using a pair correlation approximation to $S_{e}$. 
\end{abstract}
\maketitle

The behaviour of water is anomalous  when compared to simple liquids for which
the structure and dynamics is dominated by strong, essentially
isotropic, short-range repulsions \cite{ms98,pgd03}. For example, 
over certain ranges
of temperature and pressure,the density of water increases
with temperature under isobaric conditions (density anomaly) while
the self-diffusivity increases with density under isothermal conditions
(diffusional anomaly).  Experiments as
well as simulations suggest that  the anomalous thermodynamic and kinetic 
properties of water are due to the fluctuating, three-dimensional, locally
tetrahedral hydrogen-bonded network. Water-like anomalies are seen in 
other tetrahedral network-forming liquids, such as silica,
as well as  in  model liquids with isotropic core-softened or two-scale
pair potentials \cite{abhst00,ht02,vps01,ed01,sdp02,eaj99,eaj01,ybgs05,kbszs05}. 

In the case of liquids such as water and silica, a quantitative connection 
between the structure of the  tetrahedral network and the macroscopic density 
 or temperature variables can be made  by introducing
order metrics to  gauge the type as well as the extent of
structural order \cite{ed01,sdp02}. 
The local tetrahedral order parameter, $q_{tet}$, associated
with  an atom $i$ (e.g. Si atom in SiO$_2$)  is defined as
\be
q_{tet} = 1 -\frac{3}{8}\sum_{j=1}^3\sum_{k=j+1}^4 (\cos \psi_{jk}+ 1/3)^2
\ee
where $\psi_{jk}$ is the angle between the bond vectors {\bf r}$_{ij}$
and {\bf r}$_{ik}$ where $j$ and $k$ label the four nearest neighbour
atoms of the same type \cite{ed01}. 
The translational order parameter, $\tau$, measures the extent of
pair correlations present in the system and is defined as
\be \tau = (1/\xi_c ) \int_0^{\xi_c} |g(\xi ) -1|d\xi \ee
where $\xi =r\rho^{1/3}$, $r$ is the pair separation
 and $\xi_c$  is a suitably chosen cut-off distance\cite{edt03}.
Since $\tau$ increases as the random close-packing limit
is approached, it  can be  regarded as measuring the
 degree of density ordering.  At a given temperature, $q_{tet}$ will show a 
maximum and $\tau$ will show a minimum as a function of density; the loci of 
these extrema in the order define a structurally anomalous region in the
density-temperature ($\rho T$) plane.  Within this structurally anomalous
region, the tetrahedral and translational
order parameters are found to be strongly correlated. The region of the 
density anomaly,
where $(\partial \rho /\partial T)_P>0$, is bounded by the structurally
anomalous region. The diffusionally anomalous region 
($(\partial D /\partial\rho)_T
>0$) closely follows  the boundaries of the structurally anomalous 
region, specially at low temperatures. In water, the structurally
anomalous region encloses the region of anomalous diffusivity while 
this is reversed in silica. 

The pattern of nested anomalies seen in tetrahedral liquids can be 
reproduced by a model liquid with an isotropic, two-scale ramp (2SRP) pair
potential with the crucial difference that $q_{tet}$ 
must be replaced by an icosahedral order parameter, $q_{ico}$,
which is defined for a particle $i$ as \cite{snr83}
\be q_{ico} = \left[ \frac{4\pi}{2l+1} \sum^{m=l}_{m=-l}|\bar{Y}_{lm}|^2 \right]^{1/2} \ee
where $\bar{Y}_{lm}$($\theta$,$\phi$) denotes the spherical harmonics of 
order $l=6$ averaged over bonds connecting  particle $i$ with its 12 nearest 
neighbours.  The 2SRP potential has two length scales:
the  hard-core and soft-core diameters,  $\sigma_0$ and $\sigma_1$ respectively
with $\sigma_0/\sigma_1 =0.568$. A linear  repulsive ramp potential, $u(r) 
=(U_1/\sigma_1) (\sigma_1 -r)$ acts when the interparticle
distance $r$ lies  between $\sigma_0$ and $\sigma_1$. At low temperatures,
the 2SRP liquid shows a maximum in $q_{ico}$ and a minimum in $\tau$
as a function of  $\rho$, resulting in a well-defined structurally anomalous
region.

Based on simulations of silica and the two-scale ramp(2SRP) liquid, here
we show that the excess entropy, $S_{e}$, is the critical 
quantity connecting the structural, diffusional and density anomalies in 
water-like
liquids where $S_{e}$ is the difference between 
the entropy of the liquid and that of the ideal gas under identical density 
and temperature conditions. 

Molecular dynamics (MD) simulations  in the canonical (NVT)
ensemble have been performed for the liquid phase of the BKS model of silica
\cite{vps01,bks90} for which computational details are given in 
ref.\cite{smc06}. NVT-MD simulations for the Lennard-Jones (LJ) and 
NVT Monte Carlo simulations for the 2SRP liquid were performed using
a 256-particle cubic simulation cell. Excess entropy for these systems
was  estimated to better than 90\% accuracy using the pair correlation
contribution, $S_2$ \cite{cc06,be89,lh92,jz05}.
For  a one-component liquids with  isotropic pair interactions,
$S_2/Nk_B = -2\pi\rho \int \left[ g(r) \ln g(r) - g(r) +1 \right] r^2 dr$ 
where $g(r)$ is the pair correlation function\cite{be89}.  For a binary liquid 
mixture of two species A and B interacting via
isotropic potentials, this expression can be suitably generalised 
\cite{lh92} and applied to  BKS silica to give a total
entropy, $S= S_{id} + S_2$ which is consistently 6.5$\pm$0.3\% higher than the 
values  in ref.\cite{vps01} for the 3.0 g cm$^{-3}$ isochore.

We first consider the connection between $S_2$ and
the density anomaly.  Figure 1 shows the behaviour of the excess entropy, 
$S_2$, and the 
configurational energy, $\expt{U}$, as a function of density
for silica and for the two-scale ramp liquid. The $S_2 (\rho )$ curves
of the two systems are qualitatively the same even though the
dependence of $\expt{U}$ on $\rho$ is completely different.
The non-monotonic behaviour of $S_2$ with  a well-defined maximum and minimum 
results in extrema in the total entropy, $S = S_{id} + S_{e}$, since
$S_{id}$ is a monotonic function of density along an isotherm.
At these extrema, $(\partial S/\partial \rho )_T = V^2(\alpha /\kappa )$
where $\alpha$ is the isobaric expansion coefficient and $\kappa$ is the
isothermal  compressibility. Provided the liquid is mechanically stable
($\kappa > 0$), maxima and minima in $S$ must correspond to state points
which lie at the boundaries of the region of the density anomaly where
$\alpha$ equals zero.

A relationship between $S_{e}$ and a suitably scaled diffusivity, $D^*$,
is expected on the basis of corresponding states relationships of
the form $D=A\exp (\alpha S_{e})$ for dense liquids 
\cite{yr99,md96,has00,sag01}. We have scaled diffusivity for the binary
SiO$_2$ system using an Enskog-type scaling \cite{has00}
  and the resulting $D^*(\rho )$ curves have maxima and minima 
who locations coincide with those of $S_2(\rho )$ (cf. Figs. 1(a) and 2(a)).
The 2SRP system, for which $D(\rho )$ data is given in 
refs.\cite{kbszs05,ybgs05}, reveals the same correlation between $S_2$ and
$D$. Figure 2(b) shows the strong linear dependence of $\ln D^*$ 
on $S_2$ with a slope of 1.5 for liquid silica.  Consequently, the 
non-monotonic behaviour of $S_2$ as a function of $\rho$ at 
low temperatures should give rise to a region of anomalous diffusivity
in the $\rho T$-plane, even if the underlying diffusional mechanisms 
in simple and tetrahedral liquids are widely different \cite{mc04,mcr05}.

In the case of supercooled water, the configurational
entropy ($S_{conf}$), defined as the number
of inherent structure basins accessible to the system,  has been shown to
mirror the variation in diffusivity ($\ln D$) with temperature in
an analogous manner to the $S_{e}$ variation discussed above \cite{sslss00}. 
$S_{e}$ is a more convenient quantity to deal with it since it can be
readily obtained from calorimetric data or simulations and is well
approximated by a structural measure such as $S_2$.
While $S_{e}$ can be defined for any system,
$S_{conf}$ can be unambiguously defined only  for supercooled liquids where
there is a clear separation of time-scales between interbasin and 
intrabasin motions.

From the definitions of  $S_2$ and  $\tau$, both of which depend on
deviations of $g(r)$ from unity, it is to be expected 
that variations in $S_2$ and $\tau$ are anticorrelated
(cf. Figs. 1(c) and 3(a)).  Thus, the existence of a maximum in
$S_2(\rho )$ at high densities implies a minimum in $\tau$ as a function
of density and the locus of such minima in the $\rho T$-plane is 
sufficient to determine the high-temperature boundary of
the structurally anomalous region.  It is interesting to consider the 
structural changes that give rise to qualitatively similar behaviour
in $\tau (\rho )$ (or $S_2(\rho )$) curves in systems as dissimilar
as silica and the 2SRP liquids. In the low $\rho$ and 
low $T$ regime for silica and other tetrahedral liquids, increasing  $\rho$
tends  to destroy local tetrahedral order thereby reducing $q_{tet}$ and
 destructure $g(r)$ thereby reducing $\tau$ and enhancing $S_2$. Since
  the  system  is very far 
from the close-packing limit,  density-induced ordering is 
negligible.  Only after reaching a  critical density when excluded volume
effects  dominate over local tetrahedral order, does further  increase in 
density  promote pair correlations and  lead to a 
decrease in $S_2$ with increasing $\rho$.  In the case of the 2SRP
liquid, there is a single length scale at very low  densities,
corresponding to the soft wall diameter, $\sigma_1 =1 $, as shown
by the $g(r)$ curve at $\rho =1.0$ in Figure 3(b). 
As density increases the second length scale $\sigma_0$ emerges.
This leads to a  peak at $r=0.568$ and the greater structure in $g(r)$
tends to  decrease $S_2$. Simultaneously,  the introduction of 
additional distances for second and higher-order neighbours implies that the
$g(r)$ function becomes less structured and plateau-like at intermediate
distances. The net effect is a maximum in $\tau$ at $\rho =1.3$, followed by a decrease until a minimum value is reached at $\rho$=1.9.
Further increase in $\rho$ leads to a more structured $g(r)$ as
the second short length scale ($\sigma_0$) becomes dominant  and $\tau$ begins
to increase.  

We now consider the chain of arguments that relates  $S_2$ and local 
orientational order, $q$.  $S_2$ and  $\tau$  contain
the same structural information; therefore it is sufficient to
consider  the relationship between $\tau$ and $q_{tet}$ or $q_{ico}$. 
 The characteristic feature of the water-like
liquids studied here is that the correlation between $\tau$ and $q$
breaks down with density because the energetically
favourable local structures have very different symmetries in the
low density and high density limits. 
Systems, such as LJ, characterized by strong, short-range isotropic repulsions
have local icosahedral order in the liquid phase and hcp or fcc
packing in the solid phase. Increasing density enhances particle correlations
and local structure leading to a strong correlation between  
$\tau$ and $q_{ico}$ \cite{edt03}. In contrast, the 
stable crystalline forms of the 2SRP system at
low and high densities are fcc and rhombohedral respectively \cite{eaj01} 
suggesting that
the low density liquid will have local, icosahedral order which will be
attenuated with increasing density; consequently the correlation between
$q_{ico}$ and $\tau$ will break down with density.  In the case of tetrahedral
liquids, the local geometry at low and high densities is tetrahedral and
icosahedral respectively and therefore there is a density-driven breakdown
of the correlation between $q_{tet}$ and $\tau$. 
Figure 4 illustrates how
$S_2$ shifts from being negatively correlated with $q_{tet}/q_{ico}$ in the 
anomalous regime to positively correlated  in the 
normal regime along various
isotherms as a consequence of the changing relationship between
translational and orientational order. 

It is useful to apply the above arguments to the one-scale ramp (1SRP) liquid
with $\sigma_0=0$ which displays density and
diffusional anomalies but not the structural anomaly since
$q_{ico}(\rho )$ does not have a maximum \cite{ybgs05}. The 2SRP and 1SRP 
liquids show  the same qualitative behaviour in $\tau$ and $S_2$ 
with an effective second length scale originating from a peak in $g(r)$ close 
to zero at intermediate and high densities. However, there is no
density-induced shift in symmetry of local order since 
the low and high density crystalline
structures are fcc and hcp respectively, both of which are compatible with
icosahedral local order in the liquid. Consequently, 1SRP does not display
a well-defined structural anomaly.

We now consider the relationship between the positional or configurational
contribution to the entropy ($S_{e}$) and internal energy ($\expt{U}$).
Figure 5 shows the striking contrast between the $S_{e}$ versus $\expt{U}$ 
plots for silica, 2SRP and Lennard-Jones (LJ) liquids, generated
by varying density along different isotherms. The LJ system
shows a strong positive correlation between $S_2$ and $\expt{U}$.
In the case of the 2SRP liquid, the positive correlation between $S_2$ and
$\expt{U}$ in the anomalous regime shifts to a negative correlation at high
densities. The $S_2$ versus $\expt{U}$ curves for the tetrahedral liquid 
show distinct low-density and high-density segments. In the anomalous regime, 
the entropy of the high
density segment always lies above that of the low-density one, while in the
normal regime this order is reversed; the two segments coincide for the
crossover temperature. Since it is straightforward to evaluate $S_2$ and
$\expt{U}$ from simulations or experimental data, such plots should provide
a convenient means to diagnose anomalous behaviour in a range of liquids, 
including ionic and intermetallic melts and complex fluids with ultra-soft
repulsions.

Our results demonstrate that if the excess entropy, $S_{e}$,
 of a liquid shows a well-defined minimum followed by a maximum as a 
function of density, then the system will have nested diffusional and
 density anomalies.  Since $S_{e}$ is
well approximated by the pair correlation entropy, $S_2$, a connection
with the order metrics can be readily formulated to show that
if the energetically favourable local geometries in the low and
high density limits have different symmetries, then a structurally
anomalous regime can be defined in terms of an orientational
local order parameter and a translational or pair-correlation order parameter.
Such anomalous liquids will show a characterstic fingerprint when excess
entropy is plotted as a function of configurational energy.
The  relationship between the excess entropy and the diffusional and 
density anomalies will be valid even if the pair correlation approximation is 
substituted by more accurate experimental or computational estimates. Provided
pair correlations play a dominant role in liquid state structure and dynamics,
as is likely in most systems of interest, the close correspondence between
$S_{e}$ and the translational order parameter  will remain and our 
conclusions regarding the structural anomaly will also be valid. 

{\bf Acknowledgements} CC thanks the Department of Science 
and Technology, New Delhi for financial support. 
RS and SNC thank Council for Scientific and Industrial Research,
New Delhi, for the award of Junior Research Fellowships.

\begin{center}
{\bf Figure Captions}
\end{center}

\begin{enumerate}
\item Dependence of  pair correlation entropy, $S_2$, and 
configurational energy, $\expt{U}$, on density, $\rho$. 
Results for silica are shown in the left panel 
with isotherms corresponding from top  to
bottom  to T = 6000K, 5500K, 5000K, 4500K and 4000K. 
The units of $\rho$, $S_2$ and $\expt{U}$ are  g cm$^{-3}$, 
J K$^{-1}$ mol$^{-1}$ and  10$^3$ kJ mol$^{-1}$ .
Results for 2SRP liquid are shown in the right panel 
with isotherms corresponding from 
top to bottom to T = 0.2, 0.109, 0.082, 0.063, 0.045, 0.036 
and 0.027.  Reduced  units of $\rho$, $S_2$ and $\expt{U}$ are  
$\sigma_1^{-3}$, $k_B$ and  $U_1$.

\item Scaled diffusivity, $D^*$, of liquid BKS silica:
(a) Dependence of $D^*$, on $\rho$ at different $T$
and (b) Correlation plot of  $D^*$ and  $S_2$.
 The scaled diffusivity, $D^*$ equals $(D_{Si}/\chi_{Si})^{x_{Si}}
(D_{O}/\chi_{O})^{x_{O}}$, where $D_a$, $x_a$ and $\chi_a$ are the
self-diffusivity, mole fraction and scaling parameter respectively 
of component $a$ in the mixture \cite{has00}. The straight line in part(b) 
corresponds to $D=0.16\exp (1.5S_2)$; the three lowest density state points
at 4000K have been excluded from the fit.

\item (a) Dependence of translational order parameter,
$\tau$, on $\rho$  for  2SRP liquids along isotherms at $T$ values given in
Figure 1.  (b) Pair correlation function, $g(r)$, at different densities
along the T=0.027 isotherm.

\item Correlation between the excess entropy, $S_2$ and (a) the tetrahedral 
order parameter, $q_{tet}$, for silica and (b) the icosahedral order
parameter, $q_{ico}$, for 2SRP liquid. Isotherms are shown in order of
increasing $T$ from left to right at $T$ values given in Figure 1.

\item Correlation plot of  excess entropy, $S_2$ and  
configurational energy, $\expt{U}$ for (a) silica (b) 2SRP and (c) 
Lennard-Jones (LJ) along different isotherms. The highest and lowest
density state points for each isotherm are marked by horizontal and
vertical arrows. Isotherms for 2SRP liquid labelled as in Figure 1.

\end{enumerate}

\includegraphics[width=8.6cm]{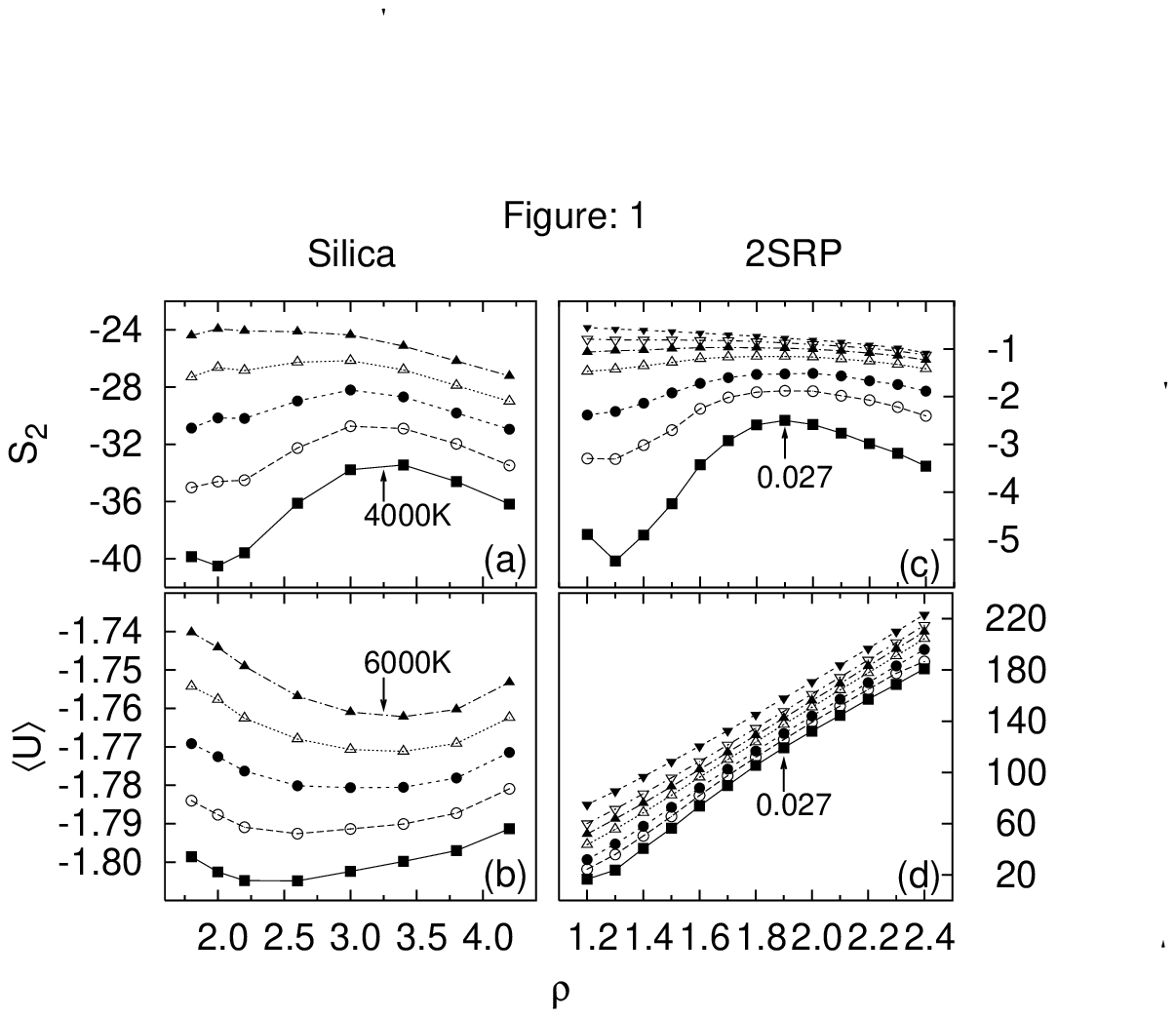}
\includegraphics[width=8.6cm]{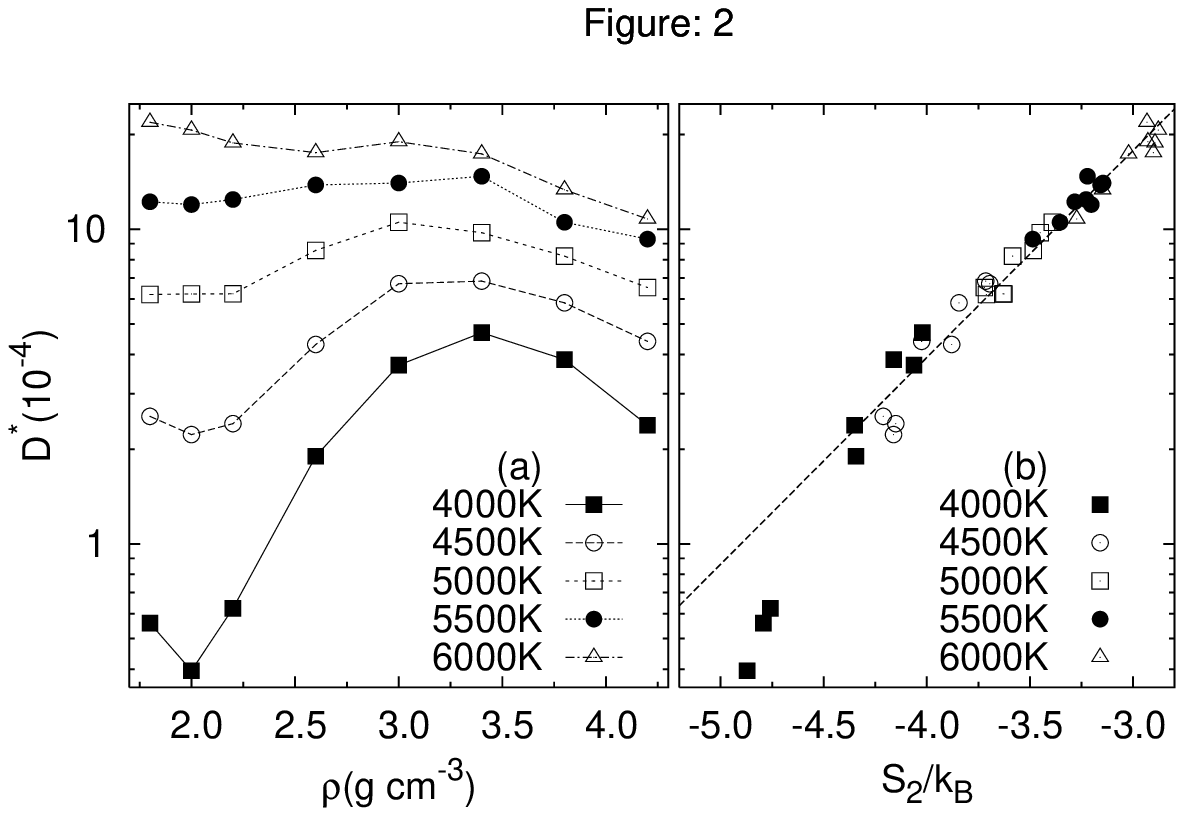}
\includegraphics[width=8.6cm]{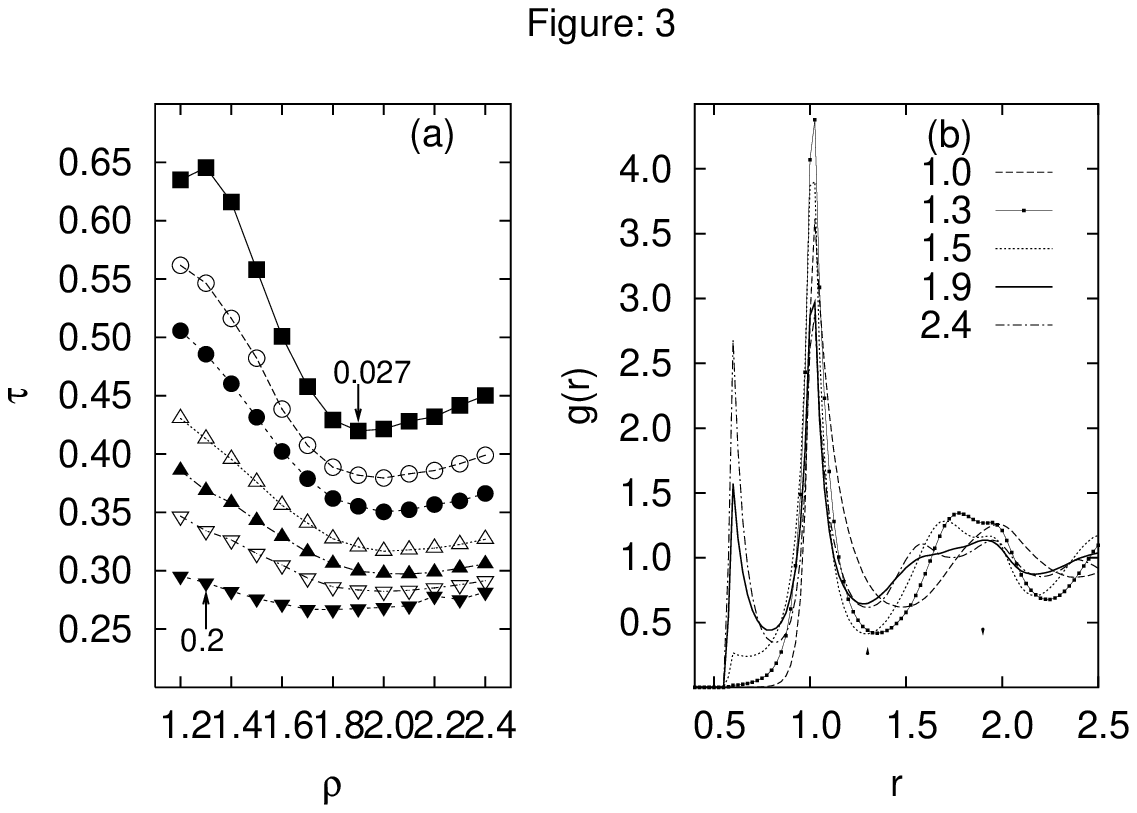}
\includegraphics[width=8.6cm]{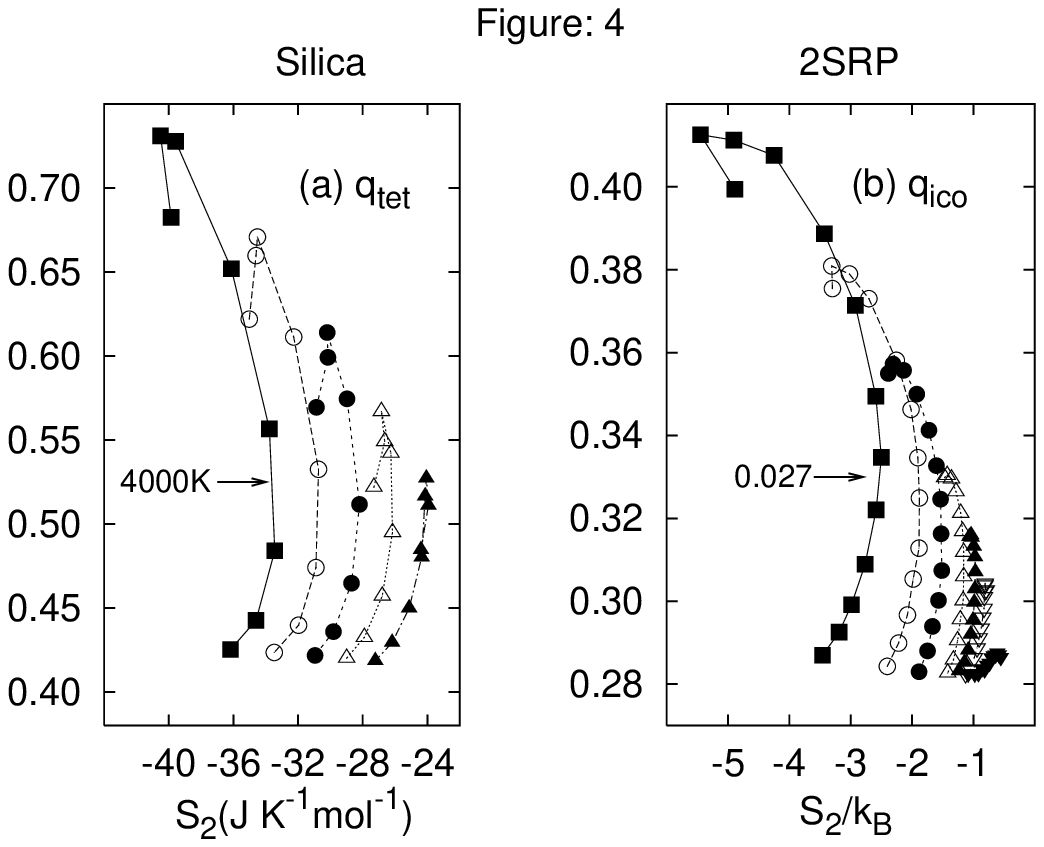}
\includegraphics[width=8.6cm]{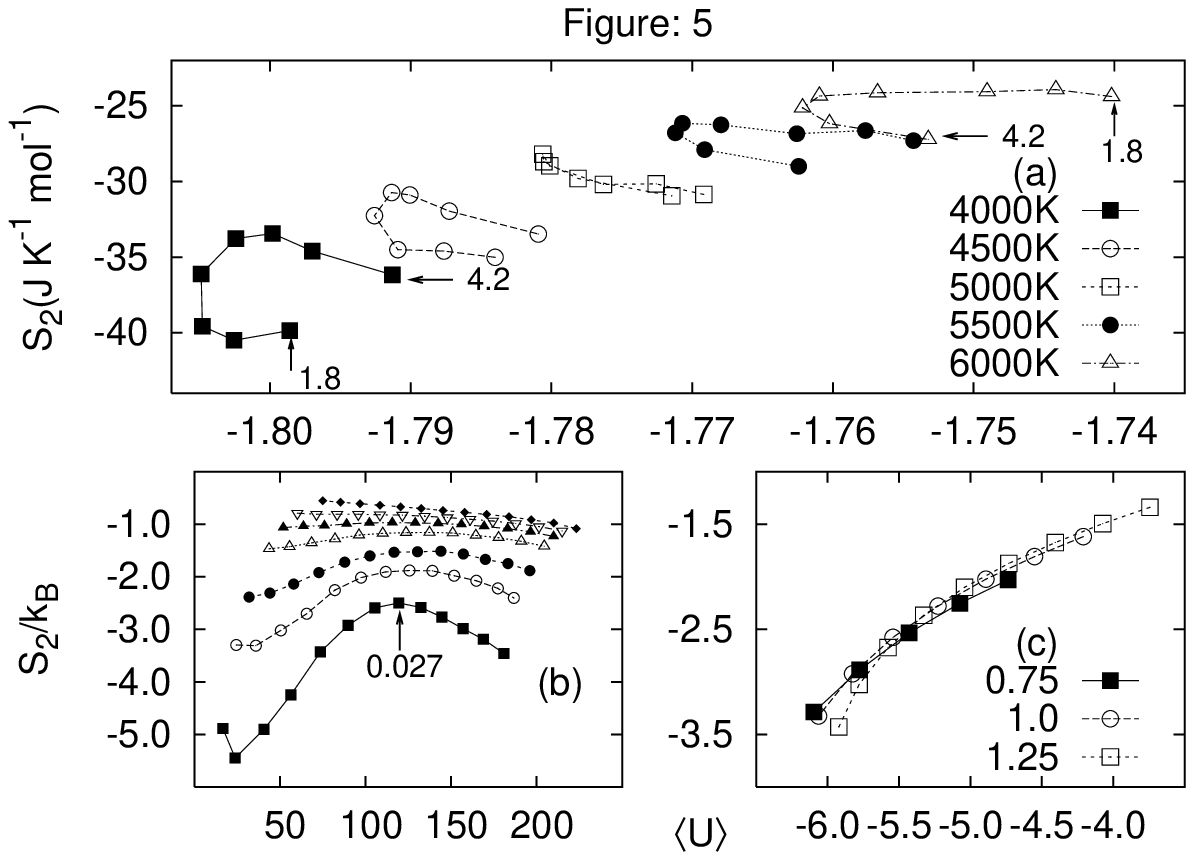}

\end{document}